\documentclass[pra,twocolumn]{revtex4}
\usepackage{graphicx}

\begin{document}
\title{Quantum dense coding in multiparticle entangled states via local measurements}
\author{Jian-Lan Chen and Le-Man Kuang\footnote{Corresponding author.}\footnote{
Email: lmkuang@hunnu.edu.cn}}
\address{Department of Physics, Hunan Normal University, Changsha 410081, China\ }

\begin{abstract}
In this paper, we study quantum dense coding between two
arbitrarily fixed particles in a $(N+2)$-particle
maximally-entangled states  through introducing an auxiliary qubit
and carrying out local measurements. It is shown that the
transmitted classical information amount through such an entangled
quantum channel usually is less than two classical bits. However
the information amount may reach two classical bits of
information, and the classical information capacity is independent
of the number of the entangled particles in the initial entangled
state under certain conditions. The results offer deeper insights
to quantum dense coding via quantum channels of multi-particle
entangled states.

\vspace{0.5cm}
 \noindent PACS number(s): 03.67.Hk, 03.65.Ud, 03.67.-a

\end{abstract}

\maketitle

Quantum dense coding (QDC) is one of the many surprising
applications of quantum entanglement. It was first proposed by
Bennett and Wiesner (BW) \cite{ben}.  QDC is a procedure in which
the transmission of classical information through a quantum
channel is enhanced by preshared entanglement between sender and
receiver.
  In the simplest example of this protocol, two people (Alice and Bob) share a pair
of entangled qubits  in a Bell state. Alice can then perform any
of the four unitary operations given by the identity $\hat{I}$ or
the Pauli matrices $\hat{\sigma}_x$, $\hat{\sigma}_y$, and
$\hat{\sigma}_z$ on her qubit. Each of these four unitary
operations map the initial state of the two qubits to a different
member of the Bell state basis. Clearly, these four orthogonal and
therefore fully distinguishable states can be used to encode two
bits of information. After encoding her qubit, Alice sends it off
to Bob who can extract these two bits of information by performing
a joint measurement on this qubit and his original qubit. This
apparent doubling of the information conveying capacity of Alice's
qubit because of prior quantum entanglement with Bob's qubit is
regarded as quantum dense coding.  QDC has been implemented
experimentally with polarization entangled photons for the case of
discrete variables \cite{mat} and with entangled states of bright
optical field produced using an Einstein-Podolsky-Rosen entangled
state for the case of continuous variables \cite{peng1,peng2}.
Some generalizations of the scheme to pairs of entangled $N$-level
systems in non-maximally entangled states \cite{liu} and to
distributed multiparticle entanglement \cite{bose} have  been
studied.  Continuous variable  QDC \cite{bra} has also been
studied widely.

 QDC between two parties can be generalized into
multiparties  \cite{bose,lee} and mixed state dense coding
\cite{bos}. Bose and coworkers \cite{bose} have generalized the BW
scheme of dense coding into multiparties in the qubit system. The
multiparty dense coding scheme can be understood in the following
way. There are $N$ users sharing an $N$-particle maximally
entangled state (MES), possessing one particle each. Suppose that
one of them, say user 1, intends to receive messages from the $N$
other users. The $N$ senders mutually decide a priori to perform
only certain unitary operations on their particles. After
performing their unitary operations, each of the $N$ senders sends
his particle to user 1. User 1 then performs a collective
measurement on the $N$ particles and identifies the state. Thus,
he can learn about the operation of each of the other $N$ users
has performed. That is to say that a single measurement is
sufficient to reveal the messages sent by all the $N$ users.

On the other hand, a controlled dense coding scheme \cite{hao} was
proposed using the three-particle Greenberger-Horne-Zeilinger
(GHZ) state. In this scheme one party (Alice) can send information
to the second party (Bob), whereas the local measurement of the
third party (Cliff) serves as quantum erasure. The quantum channel
between Alice and Bob is controlled by Cliff via the measurement
to realize controlled dense coding between Alice and Bob. In this
paper, we generalize the controlled dense coding scheme of the
three-particle GHZ  quantum channel to the case of
$(N+2)$-particle GHZ quantum channel via a series of local
measurements. Our motivation is to study quantum dense coding
between two arbitrarily fixed particles in a multiparticle MES in
order to understand how the transmitted classical information
amount through such an entangled quantum channel depends on the
number of the entangled particles.

In order to establish the notation, we first briefly review the
 QDC scheme.  QDC uses a preshared
entangled pair of qubits to transmit two bits of classical
information by sending only one qubit. It consists of three steps.
(i) An entangled state of two qubits is prepared and shared
between the two parties, say, Alice (A) and Bob (B). It can be one
of the four maximally entangled Bell states defined by
\begin{eqnarray}
\label{1} |\phi\rangle^{\pm}&=&\frac{1}{\sqrt2}(|0 0\rangle \pm |1
 1\rangle),\\
\label{2} |\psi\rangle^{\pm}&=&\frac{1}{\sqrt2}(|0 1\rangle \pm |1
0\rangle),
\end{eqnarray}
where $|i j\rangle=|i\rangle\otimes|j\rangle$ with $i, j=0,1$, and
$|0\rangle$ and $|1\rangle$ compose an orthogonal basis of each
qubit. (ii) When Bob wants to send a message to Alice, he encodes
his message on his qubit of the preshared entangled pair by
performing one of the four unitary operations $\hat{I}$,
 $\hat{\sigma}_x$, $\hat{\sigma}_y$, and $\hat{\sigma}_z$ (two bits of information).
Each of these operations transforms the entangled state uniquely
to one of the Bell states (\ref{1}) and  (\ref{2}); e.g., if the
preshared entangled state was $|\phi\rangle^{+}$, then
$\hat{I}|\phi\rangle^{+}=|\phi\rangle^{+}$,
$\hat{\sigma}_{xB}|\phi\rangle^{+}=|\psi\rangle^{+}$,
$\hat{\sigma}_{yB}|\phi\rangle^{+}=i|\psi\rangle^{-}$, and
$\hat{\sigma}_{zB}|\phi\rangle^{+}=|\phi\rangle^{-}$. Bob then
sends his qubit to Alice. (iii) Finally, Alice, the receiver,
performs a nonlocal Bell measurement to see in which of the four
Bell states the two qubits are. This reveals precisely which
operation Bob performed.

We now consider a  QDC scheme between two arbitrary fixed parties,
say Alice (user 1) and  Bob (user 2), through a quantum channel of
$(N+2)$-particle GHZ state
\begin{equation}
\label{3} |\psi\rangle=\frac{1}{\sqrt2}\left(|0 0 0 \cdots
0\rangle_{1 \cdots N+2 } + |1 1 1  \cdots 1\rangle_{1 \cdots N+2
}\right),
\end{equation}
where we assume that each user holds only one particle. This
scheme involves two basic steps: (i) prepare a two-particle
non-maximally EMS between Alice and Bob via a series of von
Neumann measurements; (ii) establish a MES between Alice and Bob
through introducing an auxiliary qubit and making a proper unitary
transformation.

Suppose that $(N+2)$-th user in the entangled state (\ref{3})
measures his qubit under the basis
\begin{eqnarray}
\label{4}
|+\rangle_{N+2}&=& \cos\theta_1|0\rangle_{N+2} + \sin\theta_1|1\rangle_{N+2}, \\
\label{5} |-\rangle_{N+2}&=& \sin\theta_1|0\rangle_{N+2} -
\cos\theta_1|1\rangle_{N+2},
\end{eqnarray}
where a measurement angle  $\theta_1$ is introduced to describe
the unitary transformation between the new basis
$\{|+\rangle_{N+2}, |-\rangle_{N+2} \}$ and the old basis
$\{|0\rangle_{N+2}, |1\rangle_{N+2} \}$.

Rewriting the GHZ state in the new basis $|+\rangle_{N+2}$ and
$|-\rangle_{N+2}$ gives
\begin{equation}
\label{6} |\psi\rangle=\frac{1}{\sqrt2}\left(|\varphi\rangle_{1
\cdots N+1}\otimes|+\rangle_{N+2} + |\phi\rangle_{1 \cdots
N+1}\otimes|-\rangle_{N+2}\right) ,
\end{equation}
where
\begin{eqnarray}
\label{7} |\varphi\rangle_{1 \cdots
N+1}&=& \cos\theta_1|0 \cdots 0\rangle_{N+1} + \sin\theta_1|1 \cdots 1\rangle_{N+1},  \\
\label{8} |\phi\rangle_{1 \cdots N+1}&=& \sin\theta_1|0 \cdots
0\rangle_{N+1} - \cos\theta_1|1 \cdots 1\rangle_{N+1}.
\end{eqnarray}

From the expression  (\ref{6}) we can see that  the von Neumann
measurement of qubit $N+2$ gives the readout $|+\rangle_{N+2}$ or
$|-\rangle_{N+2}$; each occurs with equal probability $1/2$. Now
let us analyze here the case in which $(N+2)$-th particle's
measurement gives $|+\rangle_{N+2}$ and the state of qubits $1, 2,
\cdots, N+1$ collapses to $|\varphi\rangle_{1 \cdots N+1}$; the
case of $|-\rangle_{N+2}$ can be treated in a similar way.
Generally, $|\varphi\rangle_{1 \cdots N+1}$ is not maximally
entangled and the success probability of dense coding with
$|\varphi\rangle_{1 \cdots N+1}$ is less than $1$.

 Then, $(N+1)$-th user measures his qubit under the
following basis
\begin{eqnarray}
\label{9} |+\rangle_{N+1}&=& \cos\theta_2|0\rangle_{N+1} +
\sin\theta_2|1\rangle_{N+1}, \\
\label{10}|-\rangle_{N+1}&=& \sin\theta_2|0\rangle_{N+1} -
\cos\theta_2|1\rangle_{N+1}.
\end{eqnarray}

The $(N+1)$-particle entangled state $|\varphi\rangle_{1 \cdots
N+1}$ can be expressed in terms of the new basis $|+\rangle_{N+1}$
and $|-\rangle_{N+1}$  as
\begin{equation}
\label{11} |\varphi\rangle_{1 \cdots N+1}=|\varphi\rangle_{1
\cdots N}\otimes|+\rangle_{N+1}+ |\phi\rangle_{1 \cdots
N}\otimes|-\rangle_{N+1},
\end{equation}
where we have introduced $N$-particle nonmaximally-entangled
states
\begin{eqnarray}
\label{12} |\varphi\rangle_{1 \cdots
N}&=& A_{12}|0 \cdots 0\rangle_{N} + B_{12}|1 \cdots 1\rangle_{N},\\
\label{13}|\phi\rangle_{1 \cdots N}&=&C_{12} |0 \cdots
0\rangle_{N} - D_{12}|1 \cdots 1\rangle_{N},
\end{eqnarray}
with the coefficients given by
\begin{eqnarray}
\label{14} A_{12}&=&\cos\theta_1\cos\theta_2, \hspace{0.3cm}
B_{12}=\sin\theta_1\sin\theta_2, \nonumber \\
C_{12}&=&\cos\theta_1\sin\theta_2, \hspace{0.3cm}
D_{12}=\sin\theta_1\cos\theta_2.
\end{eqnarray}

The $N$-th, $(N-1)$-th, $\cdots$, and  third  users carry out
similar local measurements. After third user makes the conditional
measurement under the basis
\begin{eqnarray}
\label{15}
|+\rangle_{3}&=& \cos\theta_{N}|0\rangle_{3} + \sin\theta_{N}|1\rangle_{3}, \\
\label{16} |-\rangle_{3}&=& \sin\theta_{N}|0\rangle_{3} -
\cos\theta_{N}|1\rangle_{3},
\end{eqnarray}
one can obtain the following two-particle nonmaximally-entangled
state between Alice and Bob
\begin{eqnarray}
\label{17} |\varphi\rangle_{1 2}&=& \prod^N_{i=1}\cos\theta_i|0
0\rangle_{1 2} + \prod^N_{i=1}\sin\theta_i|1 1\rangle_{1 2},
\end{eqnarray}
which may be used to act as a quantum channel to convey classical
information between Alice and Bob.

 In order to obtain a MES  between Alice (particle
$1$) and Bob (particle $2$), one can purify entanglement from the
two-particle nonmaximally-entangled state (\ref{17}) through
introducing an auxiliary qubit with the orthogonal basis
$|0\rangle_{aux}$ and $|1\rangle_{aux}$, and performs the
following  unitary operation on her qubit $1$ and the auxiliary
qubit
\begin{eqnarray}
\label{18}&&\hat{U}_{1  aux}=\left(
  \begin{array}{cccc}
   u_N& 0& \sqrt{1-u^2_N}& 0 \\
   0 & 1& 0& 0 \\
   0& 0& 0& -1 \\
  \sqrt{1-u^2_N} & 0& -u_N& 0
  \end{array}
 \right),
\end{eqnarray}
which is written down under the two-qubit basis
$\{|0\rangle_{1}|0\rangle_{aux},  \{|1\rangle_{1}|0\rangle_{aux},
\{|0\rangle_{1}|1\rangle_{aux},\{|1\rangle_{1}|1\rangle_{aux} \}$.
The parameter in the unitary transformation  (\ref{18}) $u_N$ is
determined by $N$ transformation angle $\theta_i$, and  defined by
\begin{equation}
\label{19} u_N=\prod^N_{i=1}\tan\theta_i.
\end{equation}

The collective unitary operation $\hat{U}_{1 aux}\otimes\hat{I}_B$
transforms the direct product  state $|\varphi\rangle_{1 2}
\otimes|0\rangle_{aux}$ to the three-particle entangled state
\begin{eqnarray}
\label{20} |out\rangle &=& \hat{U}_{1  aux}\otimes
\hat{I}_B|\varphi\rangle_{1 2}\otimes|0\rangle_{aux}  \nonumber\\
&=&\sqrt2\sin\theta|\phi^+\rangle_{12}\otimes|0 \rangle_{aux} \nonumber\\
& & + \cos\theta \sqrt{1-\tan^2\theta} |1 0\rangle_{12}\otimes|1
\rangle_{aux},
\end{eqnarray}
where $|\phi^+\rangle_{12}$ is a Bell state of particle $1$ and
particle $2$ given by Eq. (\ref{1}), $|1 0\rangle_{12}$ is the
unentangled state of the two particles,  and the effective
parameter angle $\theta$ is defined by
\begin{equation}
\label{21}\sin\theta=\frac{1}{\sqrt a}\prod^N_{i=1}\sin\theta_i,
\hspace{0.5cm}\cos\theta=\frac{1}{\sqrt
a}\prod^N_{i=1}\cos\theta_i,
\end{equation}
with
\begin{equation}
\label{22}a=\left(\prod^N_{i=1}\sin\theta_i\right)^2 +
\left(\prod^N_{i=1}\cos\theta_i\right)^2.
\end{equation}

From Eq.  (\ref{20}) we can see that a MES between Alice (particle
$1$) and Bob (particle $2$) can be obtained when that Alice
performs a von Neumann measurement on the auxiliary qubit  and has
outcome be the state $|0\rangle_{aux}$.

The average classical information amount transmitted from Alice to
Bob through using the $(N+2)$-particle MES quantum channel
(\ref{3}) and local measurements is given by
\begin{eqnarray}
\label{23}C&=&1+ 2|\sin\theta|^2 \nonumber \\
&=&1 + 2\left[1 +
\left(\prod^N_{i=1}\cot\theta_i\right)^2\right]^{-1},
\end{eqnarray}
which is the classical information capacity of the quantum
channel. Obviously, we recover the classical information capacity
of three-particle GHZ state quantum channel obtained in Ref.
\cite{hao} when $N=1$ in Eq. (\ref{23}).

 In what follows we discuss two interesting cases for the expression of  the
transmitted classical information capacity (\ref{23}). (i) The
case of $\pi/4$ transformations. In this case, one carries out
$\theta_i=\pi/4$ for all qubits except particle $1$ and $2$. From
Eq.  (\ref{23}) we can see that the classical information amount
transmitted from Alice to Bob through using the $(N+2)$-particle
MES quantum channel (\ref{3}) reaches its maximal value $2$. Hence
in this case two  bit classical information can be transmitted,
and the transmitted information amount is independent of the
number of the entangled particles in the initial entangled state
(\ref{1}). (ii) The case of $|\sin\theta_i|<|\cos\theta_i|$. In
this case, the classical information amount transmitted from Alice
to Bob through using the $(N+2)$-particle MES quantum channel
(\ref{3}) is less than two bits. From Eq.  (\ref{23}) we can see
that the more the number of the particles in the initial entangled
state is, the less the average transmitted information amount. In
particular, when the number of the particles in the initial MES
quantum channel (\ref{3}) approaches the infinity, the transmitted
information amount is equal to one bit. In this case, the
classical information capacity transmitted from Alice to Bob
through using the $(N+2)$-particle MES quantum channel (\ref{3})
can not be approved comparing with the classical communication
capacity between Alice and Bob.

In conclusion we have studied   QDC scheme between two arbitrarily
fixed particles in a $(N+2)$-particle MES through introducing an
auxiliary qubit and carrying out conditional measurements. This
scheme is an extension of the controlled quantum dense coding
scheme in Ref. \cite{hao} to the arbitrary qubit GHZ state quantum
channel. We have found that the transmitted classical information
capacity can be controlled through adjusting local measurement
angles, and usually the transmitted classical information amount
is less than two bits. It is interesting to note that the
information amount may reach the maximal value of two bits, and it
is independent of the number of the entangled particles in the
initial entangled state under certain conditions through properly
choosing conditional measurement angles. Actually, this
independence of the number of the entangled particles in the
initial entangled state reveals a kind of localization phenomenon
of the classical information capacity in the multi-particle
entangled quantum channel, this localization can not be involved
in any three-particle entangled quantum channel. We hope that this
new kind of localization phenomenon can be useful for further
studies of quantum dense coding.

\acknowledgments This work is supported by the National
Fundamental Research Program Grant No. 2001CB309310, the National
Natural Science Foundation Grant Nos. 90203018 and 10075018, the
State Education Ministry of China, and the Educational Committee
of Human Province.

\end{document}